\documentclass{article}
\usepackage{spconf,amsmath,graphicx}

\usepackage{caption}
\usepackage{wrapfig}
\usepackage{setspace}
\usepackage{cleveref}
\usepackage{enumitem}

\usepackage{multirow}
\usepackage{multicol}
\usepackage{xcolor}
\usepackage{esvect}
\usepackage{cite}

\makeatletter
\renewcommand{\section}{\@startsection
  {section}%
  {1}%
  {}%
  {-0.5\baselineskip}%
  {0.2\baselineskip}%
  {}}%

\renewcommand{\subsection}{\@startsection
  {subsection}%
  {2}%
  {}%
  {-0.1\baselineskip}%
  {0.1\baselineskip}%
  {}}%

\renewcommand{\subsubsection}{\@startsection
  {subsubsection}%
  {3}%
  {}%
  {-0.2\baselineskip}%
  {0.2\baselineskip}%
  {}}%

\g@addto@macro\normalsize{%
  \setlength\abovedisplayskip{5pt plus 2pt minus 2pt}
  \setlength\belowdisplayskip{5pt plus 2pt minus 2pt}
  \setlength\abovedisplayshortskip{4pt plus 2pt minus 2pt}
  \setlength\belowdisplayshortskip{4pt plus 2pt minus 2pt}
}

\makeatother

\Crefname{equation}{Eq.}{Eqs.}
\Crefname{figure}{Fig.}{Figs.}
\Crefname{tabular}{Tab.}{Tabs.}
\Crefname{table}{Tab.}{Tabs.}
\Crefname{section}{Sec.}{Sec.}

\def\L{{\cal L}}

\def\R{{\cal R}}

\title{Enhancing and Adversarial: Improve ASR with Speaker Labels}
\name{
\parbox{\linewidth}{\centering
Wei Zhou$^{1,2}$, Haotian Wu$^{1}$, Jingjing Xu$^{1}$, Mohammad Zeineldeen$^{1,2}$, Christoph L\"uscher$^{1,2}$,\\
Ralf Schl\"uter$^{1,2}$, Hermann Ney$^{1,2}$
\vspace{-3mm}
}}
\address{
$^1$Human Language Technology and Pattern Recognition, Computer Science Department,\\
  RWTH Aachen University, 52074 Aachen, Germany \\
$^2$AppTek GmbH, 52062 Aachen, Germany
\vspace{-2.5mm}
}
\begin{document}
%\ninept
%
\maketitle
\begin{abstract}
ASR can be improved by multi-task learning (MTL) with domain enhancing or domain adversarial training, which are two opposite objectives with the aim to increase/decrease domain variance towards domain-aware/agnostic ASR, respectively.
In this work, we study how to best apply these two opposite objectives with speaker labels to improve conformer-based ASR.
We also propose a novel adaptive gradient reversal layer for stable and effective adversarial training without tuning effort.
Detailed analysis and experimental verification are conducted to show the optimal positions in the ASR neural network (NN) to apply speaker enhancing and adversarial training.
We also explore their combination for further improvement, achieving the same performance as i-vectors plus adversarial training.
Our best speaker-based MTL achieves 7\% relative improvement on the Switchboard Hub5'00 set.
We also investigate the effect of such speaker-based MTL w.r.t. cleaner dataset and weaker ASR NN.
\end{abstract}
\begin{keywords}
ASR, speaker, multi-task, adversarial
\end{keywords}

\vspace{-0.5mm} 
\section{Introduction \& Related Work}
\vspace{-0.5mm} 
%The performance of automatic speech recognition (ASR) can be improved with additional domain information such as accents \cite{sun18accentAdvMTL, jain18accentEnh, yang18accentEnh} and noisy conditions \cite{shinohara16noiseAdv, serdyuk16noiseAdv}.
%Without extra effort to extract domain embeddings, these domain labels are usually applied in a multi-task learning (MTL) framework with ASR being the primary task.
%The secondary domain-related objectives mainly fall into two categories: 

Multi-task learning (MTL) with additional domain information such as accents \cite{jain18accentEnh, viglino19accentEnh, yang18accentEnh, sun18accentAdvMTL, tripathi18accent-gender-adv,  das2021accentAdv} and noisy conditions \cite{shinohara16noiseAdv, serdyuk16noiseAdv, meng17noiseAdv} can improve automatic speech recognition (ASR). The secondary domain-related objectives mainly fall into two categories: 

%Domain Aware Training ?
\textbf{Domain Enhancing Training (DET)} aims to enhance domain variance
% in some intermediate representations of 
in the ASR neural network (NN).
Although this appears contradictory to the ASR objective, it may help the model to learn better transformation towards a domain-aware speech content classification \cite{jain18accentEnh, yang18accentEnh, viglino19accentEnh}.
This is usually done via an additional domain discriminator.
% and requires coupled audio transcription and domain labels.
It shares certain similarity as integrating domain embeddings into the ASR NN, with a major difference of requiring the ASR NN to learn domain discriminative representations itself.
Note that this category was often referred to as domain MTL in some literature, but we use the term enhancing to explicitly differentiate it from the next category, which is also one type of MTL.

\textbf{Domain Adversarial Training (DAT)} aims to reduce domain variance in the ASR NN towards a domain-agnostic speech content classification for better generalization \cite{shinohara16noiseAdv, serdyuk16noiseAdv, meng17noiseAdv, tripathi18accent-gender-adv}, which is consistent with the ASR objective.
It also allows domain adaptation with un-transcribed audio from the target domain, where the knowledge transfer from the source domain is enabled by learning domain-invariant representations \cite{sun18accentAdvMTL, das2021accentAdv}.
The most common way to apply DAT is the gradient reversal layer (GRL) \cite{GRL2016}, which reverses the gradient from a subsequent domain discriminator, while other approaches such as \cite{bousmalis16DSN, Saito18ADR} can also be applied.
% disentangle/decouple/orthogonality ?

DET and DAT are opposite objectives, yet each of them is successfully applied to improve ASR. 
Thorough studies of their best utilization and possible combination are somewhat missing.
Some brief attempts to compare them \cite{sun18accentAdvMTL, tanaka22domainMTL} show that DAT is better than DET.
% accented speech \cite{sun18accentAdvMTL} and self-supervised speech representation learning \cite{tanaka22domainMTL}
For cross-domain sentiment classification, \cite{du20BERTdomainAwareAdv} proposed to apply DET to a self-supervised pretrained language model (LM) so that domain specific features can be more easily removed by subsequent DAT. % for better domain adaption.
But the effect and feasibility for ASR are not clear.

In this work, we study how to best apply DET and DAT with speaker labels for the speaker variability problem in ASR.
This yields the \textbf{speaker enhancing training} and the \textbf{speaker adversarial training}, respectively, where the latter is shown to improve ASR \cite{saon2017blstmSpkAdv, meng18spkAdv}.
We also propose a novel adaptive GRL for stable and effective adversarial training with negligible tuning effort.
To join the benefit of both speaker enhancing and adversarial training, we further investigate their combinations w.r.t. both training pipeline and positions in the ASR NN.
% which is another difference to \cite{du20BERTdomainAwareAdv}.
We use connectionist temporal classification (CTC) \cite{graves2016ctc} for the primary ASR task, and conduct detailed analysis and experiments on the Switchboard (SWB) \cite{swb} and Librispeech (LBS) \cite{libsp} corpora.
%We analyze the behavior of the ASR NN with speaker embeddings, which reveals clear indications of how to best apply the two objectives.
Our best speaker-based MTL achieves the same improvement as i-vectors \cite{ivector} plus adversarial training.
We also show some generalization and limitation of such speaker-based MTL.

\vspace{-0.5mm} 
\section{Multi-task Learning}
\begin{figure}[t]
\begin{minipage}[b]{1.0\linewidth}
  \centering
  \centerline{\includegraphics[width=6.5cm]{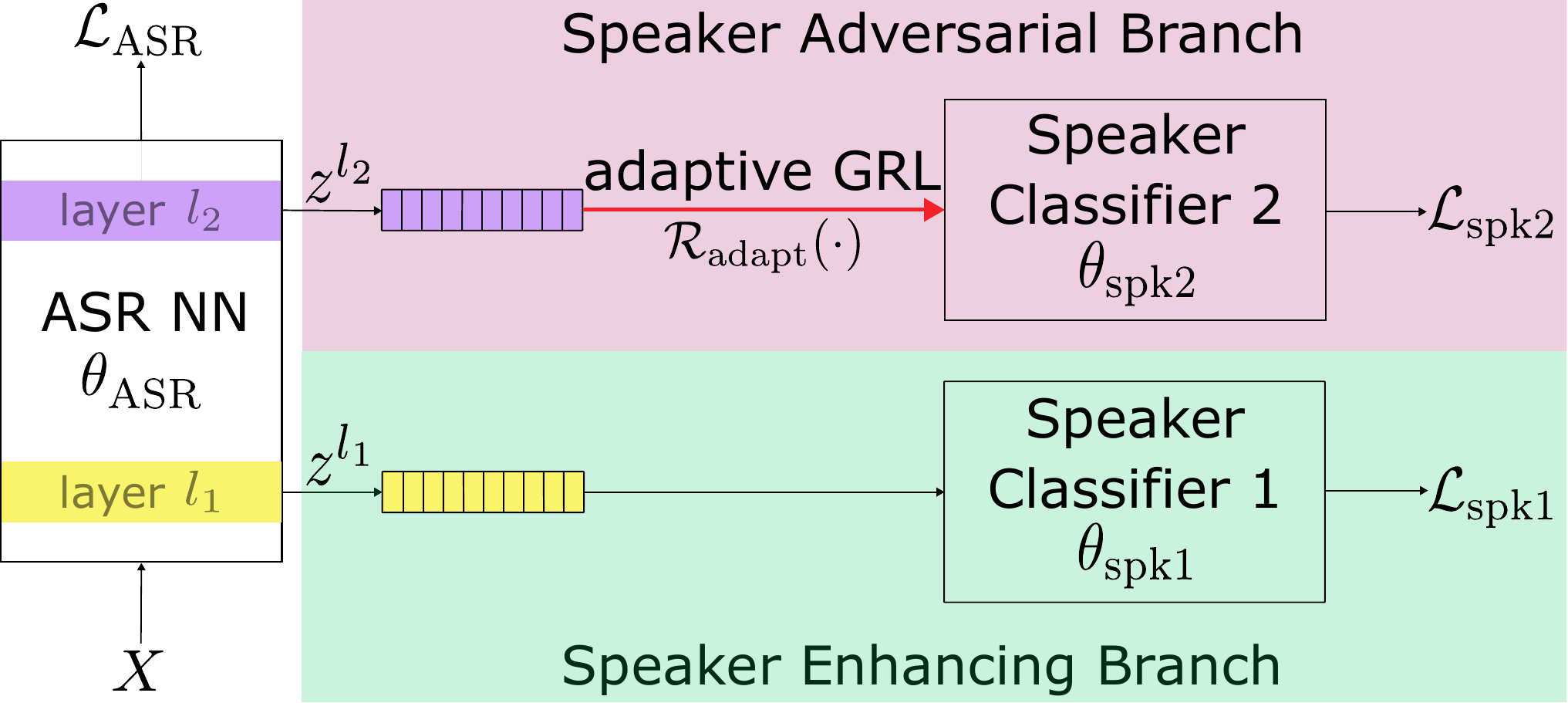}}
\end{minipage}
\caption{\it ASR + speaker-based multi-task learning framework}
\label{fig:system}
\vspace{-4mm}
\end{figure}

\vspace{-0.5mm}
With ASR being the primary task, our speaker-based MTL framework can be visualized as \Cref{fig:system}.
We denote $\theta_{\text{ASR}}$ as the parameter set of the full ASR NN, which is primarily trained with an ASR objective $\L_{\text{ASR}}$ (in this work $\L_{\text{ASR}} = \L_{\text{CTC}}$).
In the following, we define the additional speaker-based training branches in more details.

\subsection{Speaker Enhancing Training}
Let $X$ denote the acoustic feature sequence of a speech utterance, and $D_{\text{spk}}$ denote the corresponding speaker label. 
The speaker enhancing training aims to boost speaker discrimination from the output of some layer $l_1$ of the ASR NN (denoted as $z^{l_1}$).
%Although this appears contradictory to the ASR objective, it may help the model to learn better transformation towards a speaker-aware speech content classification. 
This is done via an additional speaker classifier with parameter set $\theta_{\text{spk}1}$ to maximize the target speaker posterior based on $z^{l_1}$, which yields the cross-entropy (CE) loss:\\
\scalebox{0.95}{\parbox{1.05\linewidth}{%
\begin{align*}
\L_{\text{spk}1}(\theta_{\text{spk}1}, \theta_{\text{ASR}}^{{l_1}}) = -\log P_1(D_{\text{spk}} | X; \theta_{\text{spk}1}, \theta_{\text{ASR}}^{{l_1}})
\end{align*}}}
Here we denote $\theta_{\text{ASR}}^{{l_1}}$ as the relevant part of $\theta_{\text{ASR}}$ to compute $z^{l_1}$. 
The MTL objective can then be obtained as:\\
\scalebox{0.96}{\parbox{1.05\linewidth}{%
\begin{align*}
\L_{\text{spk-enh}}(\theta_{\text{ASR}}, \theta_{\text{spk}1}) = \L_{\text{ASR}}(\theta_{\text{ASR}}) + \lambda_1 \cdot \L_{\text{spk}1}(\theta_{\text{spk}1}, \theta_{\text{ASR}}^{{l_1}})
\end{align*}}}
To avoid scale tuning for $\lambda_1$, we further apply focal loss \cite{lin2017focalloss}:\\
\scalebox{0.95}{\parbox{1.05\linewidth}{%
\begin{align*}
\lambda_1 = [ 1 - P_1(D_{\text{spk}} | X) ]^{\beta_{\text{focal}}}
\end{align*}}}
where $\beta_{\text{focal}}$ is usually omitted (set to 1).

\subsection{Speaker Adversarial Training}
The speaker adversarial training aims to force speaker-invariant representations from the output of some layer $l_2$ of the ASR NN (denoted as $z^{l_2}$). 
%The goal is towards a speaker-agnostic speech content classification for better generalization, which is consistent with the ASR objective.
Similarly, another speaker classifier with parameter set $\theta_{\text{spk}2}$ is introduced to maximize the target speaker posterior based on $z^{l_2}$.
The key difference here is the usage of the GRL $\R(\cdot)$ \cite{GRL2016}:\\
\scalebox{0.95}{\parbox{1.05\linewidth}{%
\begin{align*}
\R(z^{l_2}) = z^{l_2}, \hspace{2mm} \frac{d\hspace{0.5mm}\R}{d\hspace{0.5mm}z^{l_2}} = -\mathbf{I}
\end{align*}}}
where $\mathbf{I}$ is an identity matrix.
The GRL simply passes the input in forwarding and reverses the gradient with a negative sign in backpropagation. 
The speaker adversarial training is then achieved by applying $\R(\cdot)$ with the following CE loss:\\
\scalebox{0.95}{\parbox{1.05\linewidth}{%
\begin{align*}
\L_{\text{spk}2}(\theta_{\text{spk}2}, \R(\theta_{\text{ASR}}^{{l_2}})) = -\log P_2(D_{\text{spk}} | X; \theta_{\text{spk}2}, \theta_{\text{ASR}}^{{l_2}})
\end{align*}}}
where $\theta_{\text{ASR}}^{{l_2}}$ denotes the relevant part of $\theta_{\text{ASR}}$ to compute $z^{l_2}$. 
In this case, $\theta_{\text{spk}2}$ and $\theta_{\text{ASR}}^{{l_2}}$ lead to competing update, where the latter is reversely trained to produce confusing $z^{l_2}$ that reduces speaker discrimination.

\subsubsection{Standard Approach}
The speaker adversarial MTL loss is commonly applied as:\\
\scalebox{0.95}{\parbox{1.05\linewidth}{%
\begin{align*}
\L_{\text{spk-adv}}(\theta_{\text{ASR}}, \theta_{\text{spk}2}) = \L_{\text{ASR}}(\theta_{\text{ASR}}) + \lambda_2 \cdot \L_{\text{spk}2}(\theta_{\text{spk}2}, \R(\theta_{\text{ASR}}^{{l_2}}))
\end{align*}}}
which allows to compute all gradient updates including the competing ones in a single backpropagation.
Note that we avoid the usual min-max optimization expression here to pose a more straightforward interpretation of the GRL operation.

In this case, the same $\lambda_2 > 0$ scales the opposite update of $\theta_{\text{spk}2}$ and $\theta_{\text{ASR}}^{{l_2}}$ simultaneously, which can be difficult to tune in general.
%However, it can be very difficult in general to tune such a single scalar $\lambda_2$ for the opposite update between the domain classifier and domain adversarial primary NN.
A large $\lambda_2$ may cause unstable training (at the beginning), especially when $\theta_{\text{spk}2}$ is from-scratch initialized and produces noisy gradient.
Yet a small $\lambda_2$ may lead to insufficient adversarial training effect eventually, especially when $\theta_{\text{spk}2}$ needs to quickly catch up with (pretrained) $\theta_{\text{ASR}}$.
\cite{GRL2016} proposed to schedule an increasing $\lambda_2$ only to the GRL, and
\cite{das2021accentAdv} proposed to pretrain the domain classifier with the frozen primary NN before DAT.
Both ease the problem with some additional complexity.
But the optimal scaling still requires careful tuning and evaluation on the validation set, which can be time and resource consuming.

\subsubsection{Proposed Adaptive GRL}
\label{sec:adaptGRL}
We propose a novel adaptive GRL $\R_{\text{adapt}}(\cdot)$ for stable and effective adversarial training with negligible tuning effort:\\
\scalebox{0.95}{\parbox{1.05\linewidth}{%
\begin{align*}
\L_{\text{spk-adv}}^{\text{adapt}}(\theta_{\text{ASR}}, \theta_{\text{spk}2}) &= \L_{\text{ASR}}(\theta_{\text{ASR}}) + \L_{\text{spk}2}(\theta_{\text{spk}2}, \R_{\text{adapt}}(\theta_{\text{ASR}}^{{l_2}}))\\
\R_{\text{adapt}}(z^{l_2}) &= z^{l_2}, \hspace{2mm} \frac{d\hspace{0.5mm}\R_{\text{adapt}}}{d\hspace{0.5mm}z^{l_2}} = - \lambda_{\text{adapt}} \cdot \mathbf{I} \\
\lambda_{\text{adapt}} &= P_2(D_{\text{spk}} | X) ^{\beta_{\text{adapt}}}
\end{align*}}}
The scaling on $\L_{\text{spk}2}$ is removed to allow fast training of the discriminator $\theta_{\text{spk}2}$ all the time.
More importantly, we scale the reverse update of $\theta_{\text{ASR}}^{{l_2}}$ proportionally to the target speaker posterior from the discriminator.
Effectively, this allows more reverse update for higher speaker variance in $z^{l_2}$, and less update for lower variance, which is a desired behavior towards speaker-invariant representations.
This is the key difference to the standard approach, which does exactly the opposite.
The strength of such adaptive adversarial training can be adjusted by $\beta_{\text{adapt}}$.
We empirically find that $\beta_{\text{adapt}} = 1$ works generally well, although it can be further adjusted w.r.t. the cardinality of $D_{\text{spk}}$ and the height of $l_2$, which might affect the general magnitude of $P_2$.
Note that for a mini-batch of multiple utterances, we simply take the average $P_2$ for $\lambda_{\text{adapt}}$.

The proposed $\R_{\text{adapt}}$ also ensures a stable training.
%, even for newly initialized $\theta_{\text{spk}2}$.
At the beginning of training, a rather small $P_2(D_{\text{spk}} | X; \theta_{\text{spk}2}, \theta_{\text{ASR}}^{{l_2}})$ can be expected, which yields a rather large $\L_{\text{spk}2}$.
This leads to an aggressive training of $\theta_{\text{spk}2}$, while a very small $\lambda_{\text{adapt}}$ forbids much reverse gradient to update $\theta_{\text{ASR}}^{{l_2}}$.
After $\theta_{\text{spk}2}$ quickly catches up, an increasing $\lambda_{\text{adapt}}$ allows more reverse gradient to update $\theta_{\text{ASR}}^{{l_2}}$, which gradually reduces the speaker variance in $z^{l_2}$. 
This procedure will eventually saturate with little speaker variance left and a fluctuating $\L_{\text{spk}2}$ due to the competing update of $\theta_{\text{spk}2}$ and $\theta_{\text{ASR}}^{{l_2}}$.

% and LR schedule 
% final stability, when L_spk2 grows very large

\subsection{Enhancing + Adversarial}
\label{sec:combine}
Recent studies \cite{chang22DistilHuBERT, ashihara22deepWide} on self-supervised speech models  show that those representations from lower-middle layers are more suitable for speaker-related tasks such as speaker identification (SID), while those from higher layers are more suitable for content-related tasks such as ASR.
Inspired by these findings, we hypothesize that this might be a natural preference of DNN-based speech models, which reveals a possible combination of the two opposite objectives. 
Specifically, we propose to apply speaker enhancing training to lower layers ($l_1$) and speaker adversarial training to higher layers ($l_2$), which yields the following MTL objective:\\
\scalebox{0.95}{\parbox{1.05\linewidth}{%
\begin{align*}
\L_{\text{spk-enh-adv}}(\theta_{\text{ASR}}, \theta_{\text{spk}1}, \theta_{\text{spk}2}) &= 
\L_{\text{ASR}}(\theta_{\text{ASR}}) + \\
\lambda_1 \cdot \L_{\text{spk}1}(\theta_{\text{spk}1}, \theta_{\text{ASR}}^{{l_1}}) &+ \L_{\text{spk}2}(\theta_{\text{spk}2}, \R_{\text{adapt}}(\theta_{\text{ASR}}^{{l_2}}))
\end{align*}}}
The goal is to enhance speaker awareness at lower layers and 
remove residual speaker variance at higher layers, both of which may jointly benefit the primary ASR task.

\begin{figure}[t]
\begin{minipage}[b]{1.0\linewidth}
  \centering
  \centerline{\includegraphics[width=8cm, height=3cm]{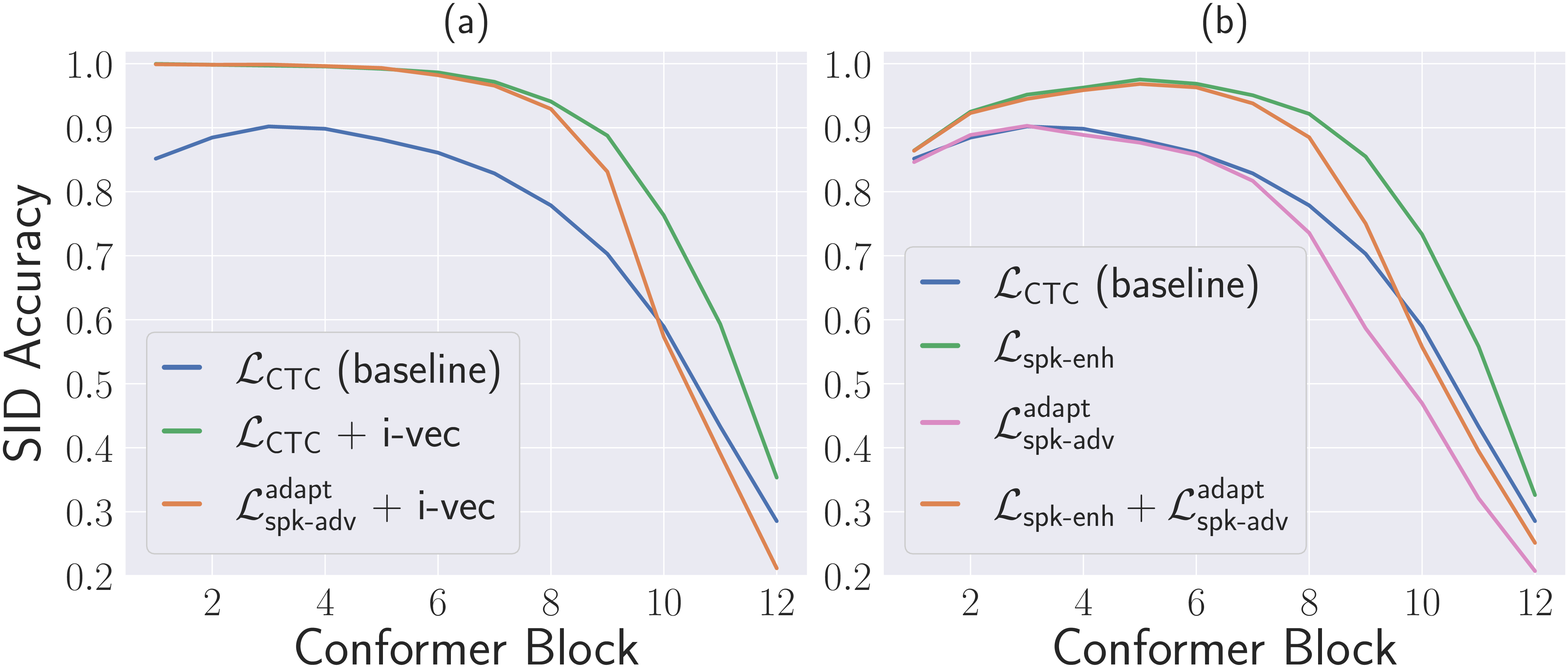}}
  \vspace{-1mm}
  \caption{\it Speaker identification (SID) accuracy analysis of each conformer block of various SWB CTC models}
\label{fig:swbsid}
\end{minipage}

\vspace{-6.7mm}
\end{figure}

\vspace{-1mm}
\section{Experiments}
\vspace{-1mm}
\subsection{Setups}
\label{sec:setup}
We conduct most experiments on the 300h SWB corpus \cite{swb} containing 520 speakers, and perform some generalization investigation on the 960h LBS corpus \cite{libsp} containing 2338 speakers.
For ASR, we use a phoneme-based CTC model, which contains a VGG network followed by 12 conformer \cite{Gulati20conformer} blocks and a softmax layer (details see \cite{zhou2022transducerTrain}).
All speaker classifiers adopt a simple MLP attention-based pooling followed by a softmax layer for the speaker posterior.
With only 1 hidden layer of 512 units for the MLP attention, each speaker classifier contains only 0.5M parameters.
For $z^{l_2}$, we always use the conformer block output, while for $z^{l_1}$, we empirically find that using the conformer block output before layer normalization gives better results.
Both $\beta_{\text{focal}}$ and $\beta_{\text{adapt}}$ are set to 1.
We use gammatone features \cite{schluter2007gt} and specaugment \cite{zoph2019specaugment} in all training.
For efficiency, we always start from a seed CTC model and reset learning rate (LR) with a constant + linear decay policy to continue training with various objectives (SWB: 25-30 epochs; LBS: 10-12 epochs).
In this case, the speaker classifiers are always from-scratch initialized.
For fair comparison, the baseline model is also continue-trained with $\L_{\text{CTC}}$ only.
By default, we apply Viterbi decoding with a 4gram word-level LM.
For SWB, we use Hub5'00 and Hub5'01 as dev and test sets, respectively.

\begin{table}[t!]
\caption{\it WER [\%] results of various CTC models on Hub5'00 including the Switchboard (SWB) and CallHome (CH) subsets, decoded with a 4gram LM.}
\label{tab:swbwer}
\centering
\scalebox{0.85}{\parbox{1\linewidth}{%
\begin{tabular}{|c|c|c|c|c|c|}
\hline
Model & \multicolumn{2}{c|}{Conformer Block} & \multicolumn{3}{c|}{Hub5'00} \\
Train & { \hspace{2mm} $l_1$ \hspace{2mm} } & $l_2$ & SWB & CH & $\sum$ \\ \hline
$\L_{\text{CTC}}$ (baseline) & \multicolumn{2}{c|}{\multirow{2}{*}{-}} & 7.6 & 17.3 & 12.4 \\
$\L_{\text{CTC}}$ + i-vec & \multicolumn{2}{c|}{} & 7.7 & 16.1 & 11.9 \\ \hline
\multirow{4}{*}{$\L_{\text{spk-adv}}^{\text{adapt}}$ + i-vec} & \multirow{4}{*}{-} & 9 & 7.5 & 16.0 & 11.8\\
& & \bf10 & 7.4 & 15.8 & \bf11.6 \\
& & 11 & 7.6 & 16.4 & 12.0  \\
& & 12 & 7.6 & 16.1 & 11.9 \\ \hline 
\hline
\multirow{7}{*}{$\L_{\text{spk-enh}}$} & 1 & \multirow{7}{*}{-} & 7.8 & 16.7 & 12.3 \\
& 2 & & 7.6 & 16.6 & 12.1 \\
& 3 & & 7.6 & 16.8 & 12.2 \\
& 4 & & 7.7 & 16.5 & 12.1 \\
& \bf5 & & 7.7 & 16.2 & \bf12.0 \\
& 6 & & 7.7 & 16.6 & 12.2 \\ 
& 7 &  & 7.8 & 17.1 & 12.4 \\ \hline
% & 12 & & 7.8 & 17.5 & 12.7 \\
\hline
\multirow{6}{*}{$\L_{\text{spk-adv}}^{\text{adapt}}$} & \multirow{9}{*}{-} & 7 & 7.9 & 17.1 & 12.5 \\
& & 8 & 7.7 & 16.8 & 12.3 \\
& & \bf9 & 7.7 & 16.4 & \bf12.1 \\
& & 10 & 7.7 & 16.9 & 12.3 \\
& & 11 & 7.7 & 16.7 & 12.2 \\
& & 12 & 7.6 & 16.8 & 12.2 \\ \cline{1-1} \cline{3-6}
\multirow{3}{*}{\shortstack[c]{\hfill $\lambda_2 = 0.1$\\ $\L_{\text{spk-adv}}$, $\lambda_2 = 0.5$\\\hfill $\lambda_2 = 1.0$}} & & \multirow{3}{*}{9} & 7.8 & 16.9 & 12.4 \\
& & & 7.7 & 16.7 & 12.2 \\
 & & & 7.6 & 16.9 & 12.3 \\ \hline
\hline
\multirow{2}{*}{$\L_{\text{spk-enh-adv}}$} & \multirow{4}{*}{\bf5} & 9 & 7.7 & 16.8 & 12.3 \\ 
 & & 10 & 7.6 & 16.6 & 12.1 \\ \cline{1-1} \cline{3-6}
\multirow{2}{*}{$\L_{\text{spk-enh}} + \L_{\text{spk-adv}}^{\text{adapt}}$} &  & \bf9 & 7.5 & 15.7 & \bf11.6 \\
& & 10 & 7.5 & 16.0 & 11.8 \\ \hline
\end{tabular}}}
\vspace{-4mm}
\end{table}

\subsection{Pre-analysis with I-vectors}
Speaker embeddings such as i-vectors \cite{ivector} are known to contain rich speaker information and improve ASR performance \cite{kitza19ivec, zeineldeen2022conf-ivec}.
To have a better idea of where to apply which speaker training, we firstly try to analyze
how much speaker variance is contained at each conformer block of the ASR NN, before and after adding such speaker discriminative representations.
We analyze this by performing SID task, where we train a speaker classifier for each conformer block output of a frozen ASR model.
This is done on the SWB corpus with a random 5\% split for evaluation.
We apply the Weighted-Simple-Add method as \cite{zeineldeen2022conf-ivec} to integrate i-vectors in the 1st conformer block during the continue-training phase.

The word error rate (WER) results of the baseline and i-vectors based CTC models are shown in \Cref{tab:swbwer}, where i-vectors give 4\% relative improvement.
The corresponding SID analysis of these two models are shown in \Cref{fig:swbsid}(a), where
i-vectors increase the SID accuracy of all conformer blocks.
The most accuracy boost occurs at blocks 1-6 (almost 100\%), which gives a clear indication for speaker enhancing training.
The increment becomes less for higher blocks with lower accuracy, which matches the ASR objective.
However, the ASR objective does not explicitly force a low residual speaker variance, which can also be seen from the final block's SID accuracy.
This seems to indicate the necessity of speaker adversarial training.
%\subsubsection{I-vectors + Speaker Adversarial Training}
As a result, we apply i-vectors together with our $\L_{\text{spk-adv}}^{\text{adapt}}$ for higher conformer blocks.
The results are also shown in \Cref{tab:swbwer}, where the best $l_2=10$ further improves the performance.
We also show the SID analysis of this model in \Cref{fig:swbsid}(a), which gives a clear proof of concept for our proposal in \Cref{sec:combine}.

\subsection{Speaker Labels Only}
We then verify those findings from the above analysis on the speaker-based MTL. 
Besides the study of best utilization, the goal is also to achieve similar improvement without the use of any speaker embeddings, which usually introduce extra complexity in run time.
% and may not be applicable in some cases.

\subsubsection{Speaker Enhancing Training}
\vspace{-0.5mm}
We apply $\L_{\text{spk-enh}}$ for conformer blocks 1-7, whose results are shown in \Cref{tab:swbwer}.
All $l_1<7$ improve the baseline.
The best $l_1=5$ achieves almost the same improvement as i-vectors.
We also perform the same SID analysis on this model, which is shown in \Cref{fig:swbsid}(b).
A very similar accuracy boost pattern as i-vectors is observed for blocks $\geq 5$, although the effect is still weaker than i-vectors.
This indicates that the NN-learned speaker discriminative representations is not as powerful as those explicitly extracted speaker embeddings, which may contain additional channel information.

% block 12 degrade the performance

\subsubsection{Speaker Adversarial Training}
We then apply $\L_{\text{spk-adv}}^{\text{adapt}}$ for conformer blocks 7-12, whose results are also shown in \Cref{tab:swbwer}.
All $l_2>7$ improve the baseline, where $l_2=9$ gives the best improvement.
We did not spend much tuning effort here, but we observe that ($l_2=12$, $\beta_{\text{adapt}}=0.5$) can also achieve this result.
%Applying adversarial training on block 7 degrades the performance, which partially justifies the benefit of speaker awareness at lower layers.
For comparison, we also evaluate the standard $\L_{\text{spk-adv}}$ for $l_2=9$ and $\lambda_2 \in \{0.1, 0.5, 1.0\}$.
As shown in \Cref{tab:swbwer}, they are all worse than our proposed adaptive adversarial training.

\begin{figure}[t]
\begin{minipage}[b]{.48\linewidth}
  \centering
  \centerline{\includegraphics[width=4cm]{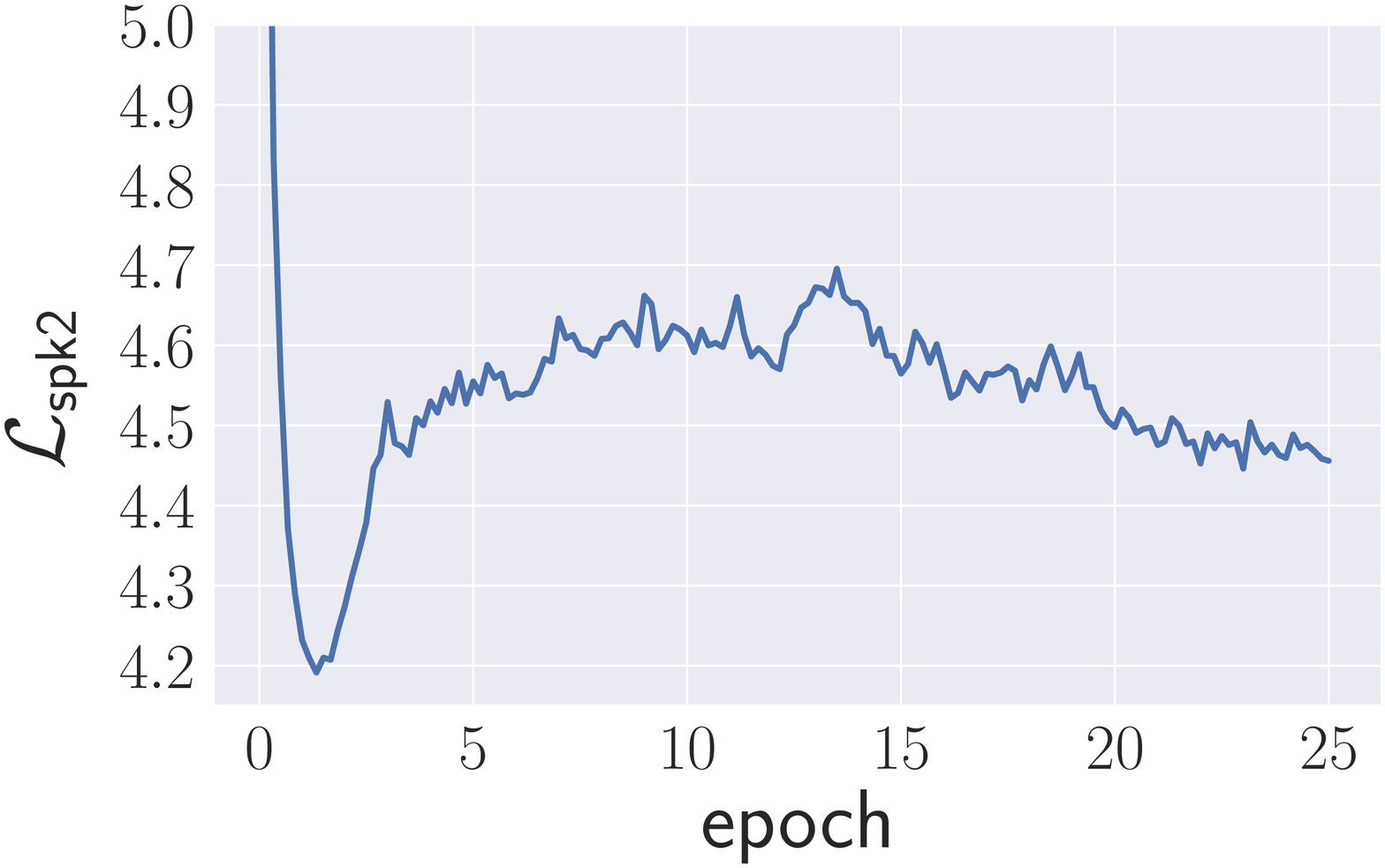}}
  \vspace{-1.3mm}
  \caption{\it $\L_{\text{spk}2}$ loss curve of the SWB CTC model trained with $\L_{\text{spk-adv}}^{\text{adapt}}$ ($l_2=9$)}
  \label{fig:advLoss}  
\end{minipage}
\hfill
\begin{minipage}[b]{0.48\linewidth}
  \centering
  \centerline{\includegraphics[width=4cm]{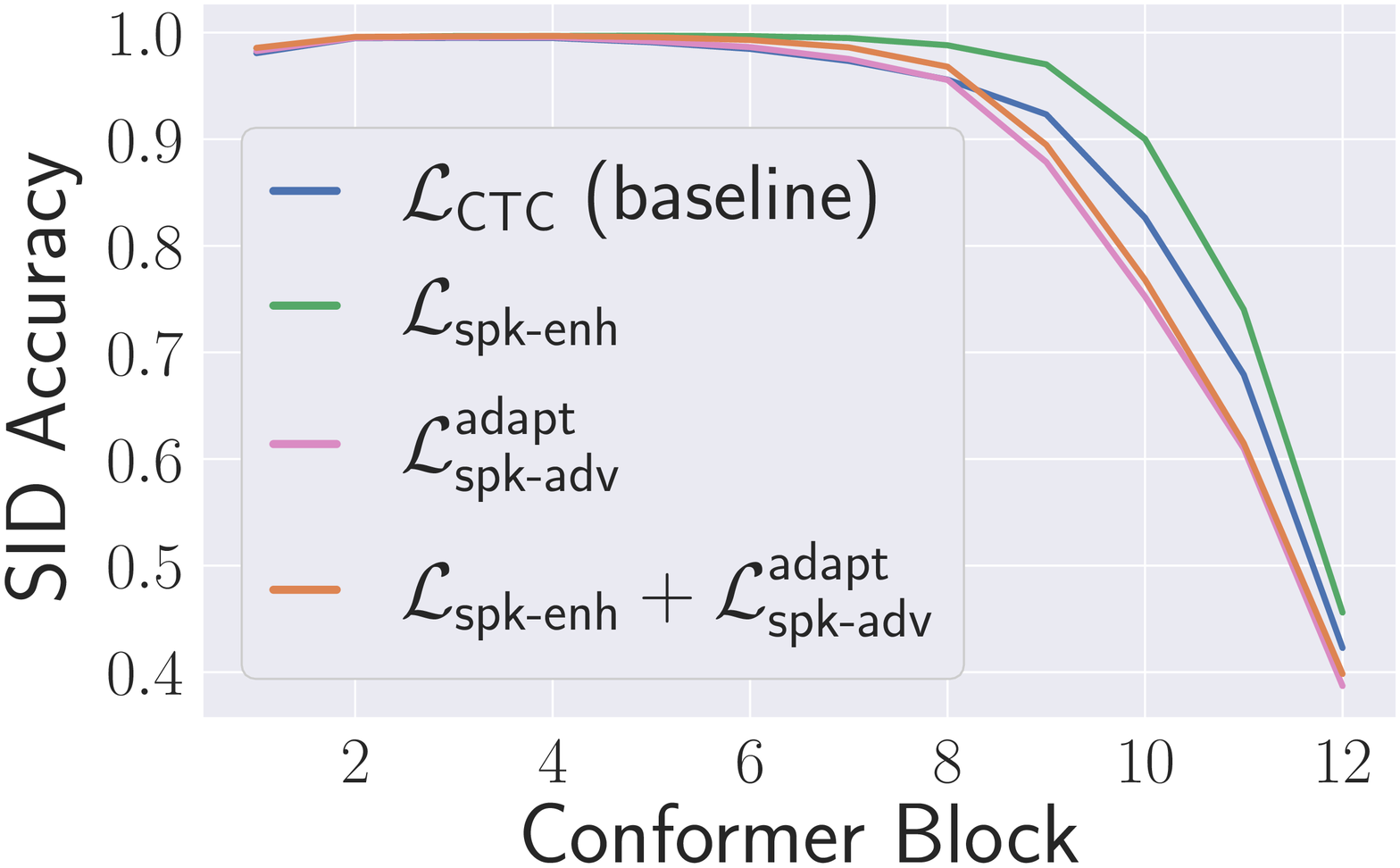}}
  \vspace{-1.3mm}
  \caption{\it SID accuracy ayalysis of each conformer block of LBS CTC models}
  \label{fig:lbssid}
\end{minipage}

\vspace{-4mm}
\end{figure}

Again, we show the best model's SID analysis in \Cref{fig:swbsid}(b), where a clear accuracy drop is observed for blocks $\geq 8$.
Together with the SID analysis for $\L_{\text{spk-adv}}^{\text{adapt}}$ + i-vectors, it seems that the adversarial training only has a local effect to reduce speaker variance of
conformer blocks $\geq l_2-1$, where lower blocks are rarely affected.
Additionally, we also show the training $\L_{\text{spk}2}$ loss curve of this model in \Cref{fig:advLoss}, which verifies the stable training behavior described in \Cref{sec:adaptGRL}.

% also useful for LR tuning

\subsubsection{Enhancing + Adversarial}
To combine the benefit of both speaker enhancing and adversarial training, we directly apply $\L_{\text{spk-enh-adv}}$ to jointly train them in one run, where the best $l_1=5$ and $l_2=9/10$ from previous experiments are used.
As shown in \Cref{tab:swbwer}, such direct joint training does not lead to further improvement over the individual best models. 
%of the two opposite objectives
%At first, we suspect a higher learning difficulty, but 
The results also do not get better after some careful tuning of LR schedule and longer training.

Based on the previous SID analysis, we suspect that the two objectives may have conflicting effect around $l_2$, as they try to increase and decrease speaker discrimination simultaneously.
As a result, we propose to sequentially apply speaker enhancing and adversarial training at lower and higher layers, respectively.
More precisely, we firstly train a speaker-aware ASR model using $\L_{\text{spk-enh}}$, and then 
fine-tune it with $\L_{\text{spk-adv}}^{\text{adapt}}$, which utilizes its local effect to gradually remove the residual speaker variance at higher layers.
We denote this as $\L_{\text{spk-enh}} + \L_{\text{spk-adv}}^{\text{adapt}}$.
As shown in \Cref{tab:swbwer}, this approach achieves further improvement, where the best ($l_1=5$, $l_2=9$) gives the same performance as the best model of $\L_{\text{spk-adv}}^{\text{adapt}}$ + i-vectors. Their SID analysis in \Cref{fig:swbsid} also show very similar pattern.

% 6.5% relative improvement over the baseline

\begin{table}[t!]
\caption{\it WER [\%] results of best SWB CTC models under various objectives and corresponding LBS CTC models under the same settings, all decoded with transformer LM.}
\label{tab:allwer}
\setlength{\tabcolsep}{0.2em}
\scalebox{0.85}{\parbox{1\linewidth}{%
\begin{tabular}{|c||c|c|c|c||c|c|c|c|}
\hline
Model & \multicolumn{3}{c|}{\small Hub5'00} & \multirow{2}{*}{{\small Hub5'01}} & \multicolumn{2}{c|}{LBS dev} & \multicolumn{2}{c|}{LBS test} \\
Train & SWB & CH & $\sum$ & & clean & other & clean & other \\ \hline
$\L_{\text{CTC}}$ (baseline) & 6.8 & 16.1 & 11.5 & 10.5 & 1.9 & 4.4 & 2.3 & 4.9 \\ \hline %cn:7.3
$\L_{\text{spk-enh}}$ & 6.6 & 15.2 & 10.9 & 10.3 & 1.9 & 4.4 & 2.3 & 4.8 \\ \hline
$\L_{\text{spk-adv}}^{\text{adapt}}$ & 6.9 & 15.1 & 11.0 & 10.4 & 1.9 & 4.3 & 2.3 & 4.8 \\ \hline
$\L_{\text{spk-enh}} + \L_{\text{spk-adv}}^{\text{adapt}}$ & \bf6.6 & \bf14.8 & \bf10.7 & \bf10.2 & \bf1.9 & \bf4.3 & \bf2.2 & \bf4.7 \\ \hline
\end{tabular}}}

\vspace{-3.5mm}
\end{table}

\subsection{Generalization and Limitation}
We evaluate the best model of each speaker-based MTL with transformer LM 1-pass decoding. As shown in \Cref{tab:allwer}, the improvements carry over to both Hub5'00 and Hub5'01 sets, where the nosier CallHome (CH) subset benefits the most.
%with more unseen speakers 
%benefits mostly (8\% relative improvement).

We then apply the best SWB settings to LBS, whose WER results and corresponding SID analysis are shown in \Cref{tab:allwer} and \Cref{fig:lbssid}, respectively.
Surprisingly, the baseline CTC model already achieves almost 100\% SID accuracy at lower conformer blocks.
%, possibly due to the cleanness of the data.
Consistent improvements are still obtained for all speaker-based MTL,
% which follows a similar pattern as SWB results.
but the overall effect is much smaller for such clean data with matched conditions.

To investigate the effect of weaker NN on speaker-based MTL, we replace conformer with a $6\times512$ bidirectional long short-term memory \cite{hochreiter1997lstm} (BLSTM) for the SWB experiments. %(LR and layers re-tuned). 
For $\L_{\text{spk-enh}}$, we get no improvement over the baseline, which indicates its strong dependency on the NN's capability to learn speaker awareness from the input $X$.
For $\L_{\text{spk-adv}}^{\text{adapt}}$, we only get minor improvement, similar as reported in \cite{saon2017blstmSpkAdv}.

% chunking does not work

%\vspace{-0.5mm}
\section{Conclusion}
%\vspace{-0.5mm}
In this work, we showed how to best apply speaker enhancing training and speaker adversarial training in a multi-task learning (MTL) framework to improve conformer-based ASR.
We proposed a novel adaptive GRL for stable and effective adversarial training without tuning effort, which outperforms the standard approach.
Our detailed analysis and experimental verification suggest to apply the enhancing training at lower layers of the ASR NN to enhance speaker awareness, and the adversarial training at higher layers to remove speaker variance matching the ASR objective. 
The two opposite objectives at different layers can also be sequentially combined for further improvement, achieving the same performance as i-vectors plus adversarial training.
This improves the baseline CTC model by 7\% relative on the Switchboard Hub5'00 set.
The effect of such speaker-based MTL is smaller on the cleaner Librispeech dataset, and is limited by weaker NN such as small BLSTM.

\vspace{-0.6mm}
\begin{center}
\bf Acknowledgements
\end{center}
\vspace{-1mm}
%\footnotesize
\scriptsize
This work was partially supported by a Google Focused Award and by NeuroSys which, as part of the initiative ``Clusters4Future'', is funded by the Federal Ministry of Education and Research BMBF (03ZU1106DA). 
The work reflects only the authors' views and none of the funding parties is responsible for any use that may be made of the information it contains.

% smaller reference
\let\normalsize\small\normalsize
% http://tex.stackexchange.com/questions/93859/condense-the-space-between-bibliographic-entries
\let\OLDthebibliography\thebibliography
\renewcommand\thebibliography[1]{
        \OLDthebibliography{#1}
        \setlength{\parskip}{-0.3pt}
        \setlength{\itemsep}{1pt plus 0.07ex}
}

\bibliographystyle{IEEEbib}
\bibliography{refs}

\begin{thebibliography}{10}

\bibitem{jain18accentEnh}
Abhinav Jain, Minali Upreti, and Preethi Jyothi,
\newblock ``{Improved Accented Speech Recognition Using Accent Embeddings and
  Multi-task Learning},''
\newblock in {\em Proc. Interspeech}, 2018, pp. 2454--2458.

\bibitem{viglino19accentEnh}
Thibault Viglino, Petr Motl{\'{\i}}cek, and Milos Cernak,
\newblock ``{End-to-End Accented Speech Recognition},''
\newblock in {\em Proc. Interspeech}, 2019, pp. 2140--2144.

\bibitem{yang18accentEnh}
Xuesong Yang, Kartik Audhkhasi, Andrew Rosenberg, Samuel Thomas, Bhuvana
  Ramabhadran, and Mark Hasegawa{-}Johnson,
\newblock ``{Joint Modeling of Accents and Acoustics for Multi-Accent Speech
  Recognition},''
\newblock in {\em Proc. {ICASSP}}, 2018, pp. 5989--5993.

\bibitem{sun18accentAdvMTL}
Sining Sun, Ching{-}Feng Yeh, Mei{-}Yuh Hwang, Mari Ostendorf, and Lei Xie,
\newblock ``{Domain Adversarial Training for Accented Speech Recognition},''
\newblock in {\em Proc. {ICASSP}}, 2018, pp. 4854--4858.

\bibitem{tripathi18accent-gender-adv}
Aditay Tripathi, Aanchan Mohan, Saket Anand, and Maneesh Singh,
\newblock ``{Adversarial Learning of Raw Speech Features for Domain Invariant
  Speech Recognition},''
\newblock in {\em Proc. {ICASSP}}, 2018, pp. 5959--5963.

\bibitem{das2021accentAdv}
Nilaksh Das, Sravan Bodapati, Monica Sunkara, Sundararajan Srinivasan, and
  Duen~Horng Chau,
\newblock ``{Best of Both Worlds: Robust Accented Speech Recognition with
  Adversarial Transfer Learning},''
\newblock in {\em Proc. Interspeech}, 2021, pp. 1314--1318.

\bibitem{shinohara16noiseAdv}
Yusuke Shinohara,
\newblock ``{Adversarial Multi-Task Learning of Deep Neural Networks for Robust
  Speech Recognition},''
\newblock in {\em Proc. Interspeech}, 2016, pp. 2369--2372.

\bibitem{serdyuk16noiseAdv}
Dmitriy Serdyuk, Kartik Audhkhasi, Philemon Brakel, Bhuvana Ramabhadran, Samuel
  Thomas, and Yoshua Bengio,
\newblock ``{Invariant Representations for Noisy Speech Recognition},''
\newblock 2016,
\newblock http://arxiv.org/abs/1612.01928.

\bibitem{meng17noiseAdv}
Zhong Meng, Zhuo Chen, Vadim Mazalov, Jinyu Li, and Yifan Gong,
\newblock ``{Unsupervised Adaptation with Domain Separation Networks for Robust
  Speech Recognition},''
\newblock in {\em {IEEE ASRU}}, 2017, pp. 214--221.

\bibitem{GRL2016}
Yaroslav Ganin, Evgeniya Ustinova, Hana Ajakan, Pascal Germain, Hugo
  Larochelle, Fran{\c{c}}ois Laviolette, Mario Marchand, and Victor~S.
  Lempitsky,
\newblock ``{Domain-Adversarial Training of Neural Networks},''
\newblock {\em Journal of Machine Learning Research}, vol. 17, pp. 59:1--59:35,
  2016.

\bibitem{bousmalis16DSN}
Konstantinos Bousmalis, George Trigeorgis, Nathan Silberman, Dilip Krishnan,
  and Dumitru Erhan,
\newblock ``{Domain Separation Networks},''
\newblock in {\em Proc. {NeurIPS}}, 2016, pp. 343--351.

\bibitem{Saito18ADR}
Kuniaki Saito, Yoshitaka Ushiku, Tatsuya Harada, and Kate Saenko,
\newblock ``{Adversarial Dropout Regularization},''
\newblock in {\em Int. Conf. on Learning Representations (ICLR)}, 2018.

\bibitem{tanaka22domainMTL}
Tomohiro Tanaka, Ryo Masumura, Hiroshi Sato, Mana Ihori, Kohei Matsuura,
  Takanori Ashihara, and Takafumi Moriya,
\newblock ``{Domain Adversarial Self-Supervised Speech Representation Learning
  for Improving Unknown Domain Downstream Tasks},''
\newblock in {\em Proc. Interspeech}, 2022, pp. 1066--1070.

\bibitem{du20BERTdomainAwareAdv}
Chunning Du, Haifeng Sun, Jingyu Wang, Qi~Qi, and Jianxin Liao,
\newblock ``{Adversarial and Domain-Aware {BERT} for Cross-Domain Sentiment
  Analysis},''
\newblock in {\em Proc. ACL}, 2020, pp. 4019--4028.

\bibitem{saon2017blstmSpkAdv}
George Saon, Gakuto Kurata, Tom Sercu, Kartik Audhkhasi, Samuel Thomas,
  Dimitrios Dimitriadis, Xiaodong Cui, Bhuvana Ramabhadran, Michael Picheny,
  Lynn{-}Li Lim, Bergul Roomi, and Phil Hall,
\newblock ``{English Conversational Telephone Speech Recognition by Humans and
  Machines},''
\newblock in {\em Proc. Interspeech}, 2017, pp. 132--136.

\bibitem{meng18spkAdv}
Zhong Meng, Jinyu Li, Zhuo Chen, Yang Zhao, Vadim Mazalov, Yifan Gong, and
  Biing{-}Hwang Juang,
\newblock ``{Speaker-Invariant Training Via Adversarial Learning},''
\newblock in {\em Proc. {ICASSP}}, 2018, pp. 5969--5973.

\bibitem{graves2016ctc}
Alex Graves, Santiago Fern{\'{a}}ndez, Faustino~J. Gomez, and J{\"{u}}rgen
  Schmidhuber,
\newblock ``{Connectionist Temporal Classification: Labelling Unsegmented
  Sequence Data with Recurrent Neural Networks},''
\newblock in {\em Proc. ICML}, 2006, pp. 369--376.

\bibitem{swb}
J.~J. {Godfrey}, E.~C. {Holliman}, and J.~{McDaniel},
\newblock ``{SWITCHBOARD: Telephone Speech Corpus for Research and
  Development},''
\newblock in {\em Proc. {ICASSP}}, 1992, vol.~1, pp. 517--520.

\bibitem{libsp}
Vassil Panayotov, Guoguo Chen, Daniel Povey, and Sanjeev Khudanpur,
\newblock ``{Librispeech: An ASR corpus based on public domain audio books},''
\newblock in {\em Proc. {ICASSP}}, 2015, pp. 5206--5210.

\bibitem{ivector}
Najim Dehak, Patrick~J. Kenny, Réda Dehak, Pierre Dumouchel, and Pierre
  Ouellet,
\newblock ``{Front-End Factor Analysis for Speaker Verification},''
\newblock {\em IEEE Transactions on Audio, Speech, and Language Processing},
  vol. 19, no. 4, pp. 788--798, 2011.

\bibitem{lin2017focalloss}
Tsung{-}Yi Lin, Priya Goyal, Ross~B. Girshick, Kaiming He, and Piotr
  Doll{\'{a}}r,
\newblock ``{Focal Loss for Dense Object Detection},''
\newblock in {\em {IEEE} International Conference on Computer Vision {ICCV}},
  2017, pp. 2999--3007.

\bibitem{chang22DistilHuBERT}
Heng{-}Jui Chang, Shu{-}Wen Yang, and Hung{-}yi Lee,
\newblock ``{Distilhubert: Speech Representation Learning by Layer-Wise
  Distillation of Hidden-Unit Bert},''
\newblock in {\em Proc. {ICASSP}}, 2022, pp. 7087--7091.

\bibitem{ashihara22deepWide}
Takanori Ashihara, Takafumi Moriya, Kohei Matsuura, and Tomohiro Tanaka,
\newblock ``{Deep versus Wide: An Analysis of Student Architectures for
  Task-Agnostic Knowledge Distillation of Self-Supervised Speech Models},''
\newblock in {\em Proc. Interspeech}, 2022, pp. 411--415.

\bibitem{Gulati20conformer}
{Anmol Gulati et al.},
\newblock ``{Conformer: Convolution-augmented Transformer for Speech
  Recognition},''
\newblock in {\em Proc. Interspeech}, 2020, pp. 5036--5040.

\bibitem{zhou2022transducerTrain}
Wei Zhou, Wilfried Michel, Ralf Schl\"uter, and Hermann Ney,
\newblock ``{Efficient Training of Neural Transducer for Speech Recognition},''
\newblock in {\em Proc. Interspeech}, 2022, pp. 2058--2062.

\bibitem{schluter2007gt}
Ralf Schl{\"u}ter, Ilja Bezrukov, Hermann Wagner, and Hermann Ney,
\newblock ``{Gammatone Features and Feature Combination for Large Vocabulary
  Speech Recognition},''
\newblock in {\em Proc. {ICASSP}}, 2007, pp. 649--652.

\bibitem{zoph2019specaugment}
Barret Zoph, Chung-Cheng Chiu, Daniel~S. Park, Ekin~Dogus Cubuk, Quoc~V. Le,
  William Chan, and Yu~Zhang,
\newblock ``{SpecAugment: A Simple Augmentation Method for Automatic Speech
  Recognition},''
\newblock in {\em Proc. Interspeech}, 2019, pp. 2613--2617.

\bibitem{kitza19ivec}
Markus Kitza, Pavel Golik, Ralf Schl{\"{u}}ter, and Hermann Ney,
\newblock ``{Cumulative Adaptation for {BLSTM} Acoustic Models},''
\newblock in {\em Proc. Interspeech}, 2019, pp. 754--758.

\bibitem{zeineldeen2022conf-ivec}
Mohammad Zeineldeen, Jingjing Xu, Christoph L\"uscher, Ralf Schl\"uter, and
  Hermann Ney,
\newblock ``{Improving the Training Recipe for a Robust Conformer-based Hybrid
  Model},''
\newblock in {\em Proc. Interspeech}, 2022, pp. 1036--1040.

\bibitem{hochreiter1997lstm}
Sepp Hochreiter and J\"{u}rgen Schmidhuber,
\newblock ``{Long Short-Term Memory},''
\newblock {\em Neural Computation}, vol. 9, no. 8, pp. 1735--1780, 1997.

\end{thebibliography}

\end{document}